\documentclass[twocolumn,showpacs,preprintnumbers,amsmath,amssymb]{revtex4}
\usepackage{graphicx}
\usepackage{dcolumn}
\usepackage{bm}
\usepackage{subfigure}
\usepackage{epsfig}
\usepackage{epsf}
\usepackage{amssymb}

\begin{document}

\title{On the Phase Diagram of QCD with Small Isospin Chemical Potential}

\author{Juliane Mirsa M\o ller}
\affiliation{%
The Niels Bohr International Academy, The Niels Bohr Institute, Blegdamsvej $17$, DK-$2100$, Copenhagen \O, Denmark.}%


\begin{abstract}
The regime of small isospin chemical potential in QCD is investigated. Using the phase quenched partition function in the $\epsilon$-regime an expression for the chiral condensate is given, which is studied in the temperature isospin chemical potential plane. Lines of constant values of the condensate are shown and it is estimated how the critical temperature varies as a function of the isospin chemical potential. Finally, the dependency of the fermion sign problem on the chemical potential and temperature is examined.
\end{abstract}

\pacs{12.39.Fe,11.15.Ha}

\maketitle
\section{\label{sec:level1}Introduction}
During the past years the phase diagram of QCD at finite baryon density has been a subject of great interest. Eventhough, a large number of phases have been suggested \cite{hands-2002-106,Frontier} only now direct confirmation by first principle calculations has emerge. One reason is that the phase of the fermion is complex, which makes standard Monte-Carlo simulations possible only for small values of the chemical potential \cite{Fodor:2001pe,deforcrand-2002-642,allton-2002-66,Miyamura:2002en,Gavai:2002kq,delia-2003-67,splittorff-2006-023}.\\
The phenomenology of heavy ion collisions and neutron stars has also been contributing to the confirmation of the phase diagrams of QCD  \cite{Rajagopal:2000wf}.
In both neutron stars and relativistic ion collisions the isospin density is different from zero. Therefore, it is of great interest to study the influence of isospin chemical potentials on the phase diagram of QCD. The main goal of this paper is thus to examine the phase diagram of QCD at nonzero temperature and isospin chemical potentials.\\
The starting point will be the work of Gasser and Leutwyler, who propounded that in the $\epsilon$-regime \cite{Verbaarschot:2000dy} the low energy partition function, which is dominated by the zero modes of the Goldstone bosons, reduces to a group integral uniquely determined by the pattern of chiral symmetry breaking \cite{Gasser Leutwyler,Gasser:1987zq,Gasser:1986vb}.
Since the lightest degrees of freedom, the pions, have zero baryon charge, the low energy effective partition function at zero temperature is not affected by the baryon chemical potential. In addition, due to the complex fermion determinant it is not possible directly to simulate lattice QCD at nonzero baryon chemical potential by probabilistic methods. By instead considering the absolute value of the fermion determinant, one gets a situation which is doable. For an even number of flavors this corresponds to the product of a fermion determinant and its complex conjugate. 
The conjugate flavors correspond to ordinary fermionic flavors with the opposite sign of the chemical potential \cite{Alford:1998sd}. A theory with one flavor and a conjugate flavor is therefore identical to a theory with two flavors at nonzero isospin chemical potential $\mu_{iso}$. For further discussions of isospin chemical potential see \cite{Kogut:2002zg,Kogut:2001id,Kogut:2001if,Kogut:2002cm}. Since the pions have nonzero isospin charge, the low energy effective partition function depends on the chemical potential. Thus, by looking at the isospin chemical potential instead of a baryon chemical potential, a situation is reached, where numerical calculations and nontrivial analytic computations are possible.\\

The organization of this paper is as follows. In section $II$ the expected features of the QCD phase diagrams for nonzero temperature, baryon and isospin chemical potentials are reviewed from general arguments. QCD is considered in the $\epsilon$-regime for which the volume of Euclidean space-time is taken such that chiral perturbation theory is valid and that the Goldstone modes associated with chiral symmetry are the dominant degrees of freedom \cite{Leutwyler Smilga}. Moreover, the Goldstone field is treated as constant. This has the advantage of allowing for exact, analytic calculations. In section $III$ the effective theory is used to derive an expression for the chiral condensate in the case of two quark flavors $N_f=2$. The expression is presented in the $(\mu_{iso},T)$-plane and from the lines of constant values of the chiral condensate it is estimated how the critical temperature drops as a function of the isospin chemical potential. In section $IV$ the fermion sign 
 problem is investigated by plotting the expectation value of the phase factor in the $(\mu,T)$-plane. Finally, in section $V$ concluding remarks are given.

\section{QCD at Nonzero Chemical Potential}
The QCD partition function at temperature $1/\beta$ and chemical potential $\mu$ is given by
\begin{equation}
\mathcal{Z}_{QCD} =\sum_k e^{-\beta(E_k-\mu)}, 
\label{eq:Hej1}
\end{equation}
where the sum is over all states. This partition function can be rewritten as the Euclidean QCD partition function
for $N_f$ quark flavors
\begin{equation}
\mathcal{Z}^{N_f}(m;\mu)=\langle \prod_{f=1}^{N_f}\det(D+\mu_f \gamma_0 + m_f) \rangle,
\label{eq:Euclidean}
\end{equation}
where the average is over the Euclidean Yang-Mills action and the Dirac operator is $D=\gamma_\mu(\partial_\mu+ iA_\mu)$, with $\gamma_\mu$ the Euclidean $\gamma$-matrices, and $A_\mu$ an $SU(N_c)$ valued gauge potential. The quark masses are denoted by $m_f$ and the chemical potential for each flavor is denoted by $\mu_f$. Below, focus will mainly be on QCD with two quark flavors and nonzero baryon number and isospin chemical potential, which in this case are defined as 

\begin{eqnarray}
\mu_B &=& \mu_1+\mu_2,\\
\mu_{iso} &=& \mu_1-\mu_2.
\label{eq:chemical potential}
\end{eqnarray}

\begin{figure}[htb]
\epsfxsize=1.6in
{\epsffile{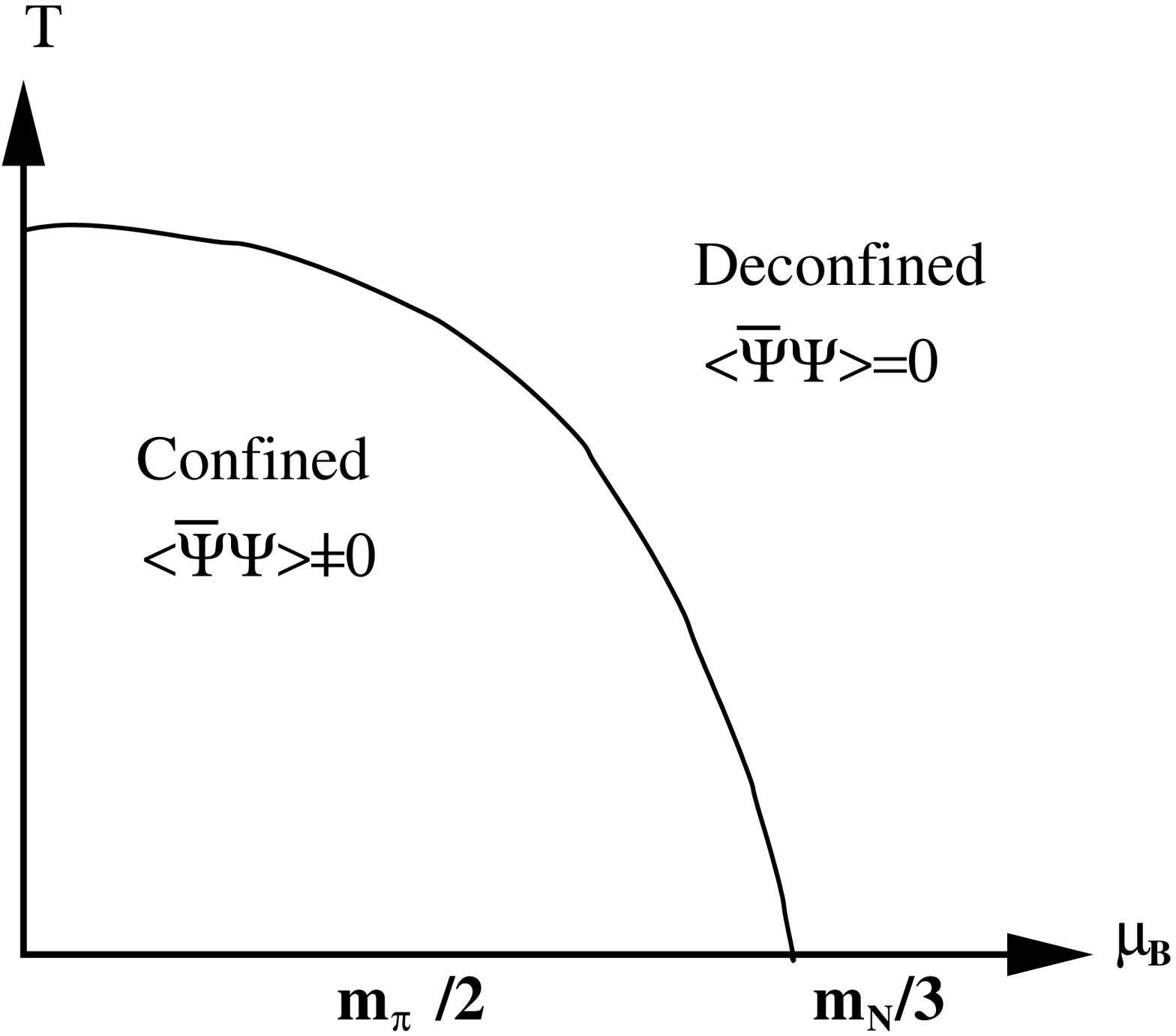}}
\epsfxsize=1.6in
\epsffile{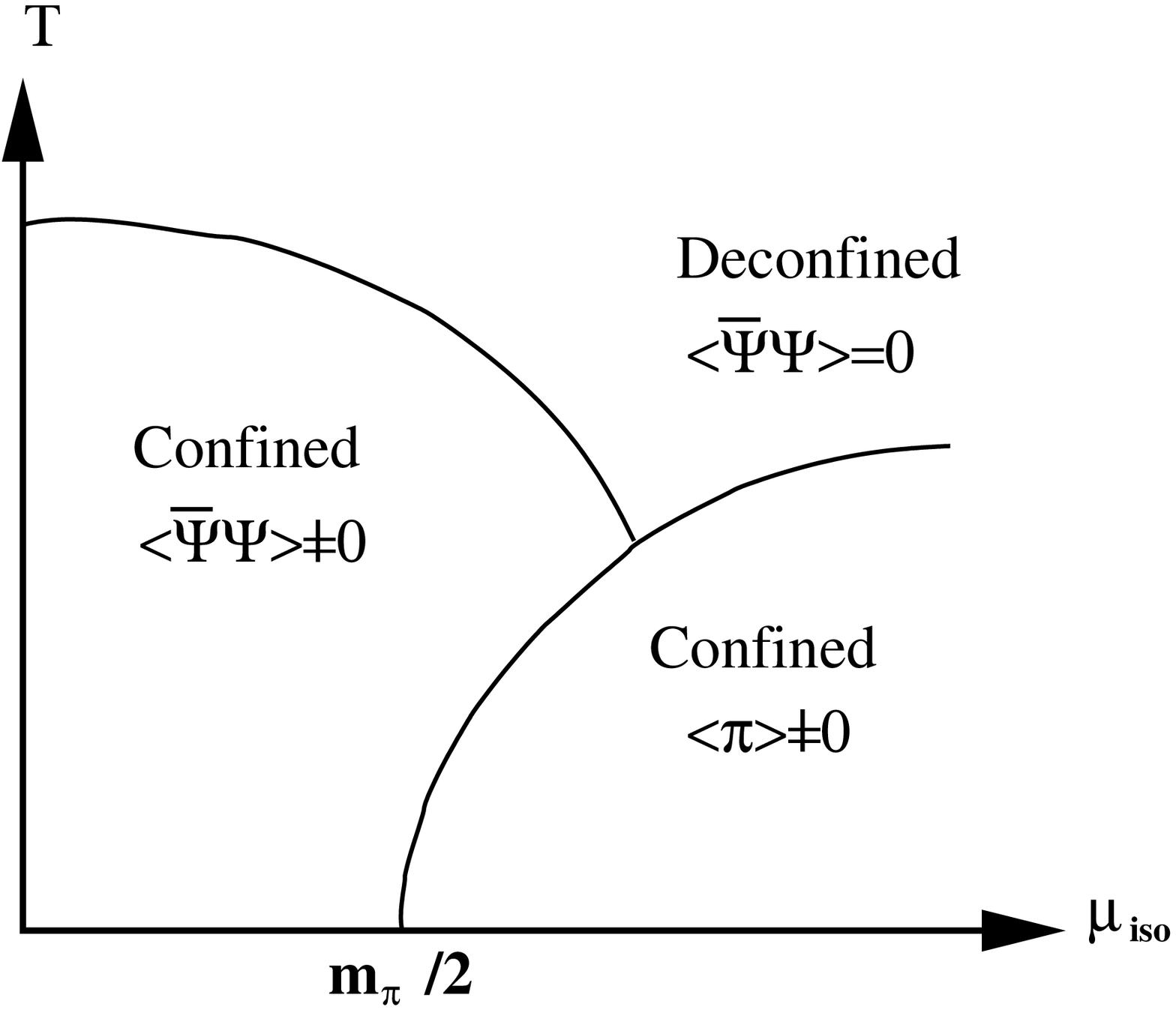}
\vspace{0.1in}
\epsfxsize=1.6in
\caption{\textit{\label{fig:Phase} A schematic phase diagram of QCD in the temperature baryon chemical potential plane (left) and the temperature isospin chemical potential plane (right). The $m_\pi$ denotes the pion mass and $m_N$ denotes the nucleon mass.}}
\end{figure}

Before proceeding to the expression for the chiral condensate some general remarks on the QCD phase diagrams are given. For a further discussion of the QCD phases see \cite{Klein:2003fy,Toublan:2004bj,deforcrand-2007-237}. Figure \ref{fig:Phase} 
shows the simplified phase diagram of QCD in the temperature baryon chemical potential plane and in the temperature isospin chemical potential plane. At temperature $T \sim 170$ MeV QCD undergoes a phase transition from a chirally broken phase to a chirally symmetric phase. Since lattice QCD requires a real action, information of the structure of the phase diagram along the chemical potential axis cannot be achieved reliably using that regularization. Therefore, most of the knowledge of QCD at nonzero baryon density is based on models and general properties of the phases of QCD. For a discussion see \cite{Halasz:1998qr}.\\
In the case of isospin chemical potential an analysis of the chiral Lagrangian to one-loop order in the low temperature range shows that the pions Bose condense for $\mu_{iso} > m_\pi /2$, where $m_\pi$ is the physical pion mass \cite{Splittorff:2001fy, Splittorff:2002xn}. Contrary to the partion function for $\mu_B \neq 0$, the partition function for $\mu_{iso}\neq 0$ can be simulated by Monte-Carlo methods. When using an isospin chemical potential instead of a baryon chemical potential the fermion determinant is real and therefore traditional numerical lattice methods can be applied \cite{kogut-2004-70}.\\
\\
Measurements of lattice QCD are necessarily performed at finite volume.
In this paper the standard choice in lattice regularization of QCD is used. This choice considers a torus, where the extension in the Euclidean time direction determines the temperature $L_4 = 1/T$ and the quantities $L_1,L_2,L_3$ specify a three dimensional box. By considering a torus with $L_1=L_2=L_3=L_4=L$ the volume of the Euclidean space-time is thus defined by $V=L^4$.\\
Moreover, the effective theory is considered in the presence of an isospin chemical potential. With the usual pattern of spontaneous chiral symmetry breaking for two light flavors, the theory is described by a Lie-group valued field $U(x) \in SU(2)$. As mentioned, focus will be on the $\epsilon$-regime, also known as the microscopic domain of QCD. In this regime the chiral Lagrangian is treated as a perturbative expansion around the zero-momentum modes in a finite volume.

\section{Effective Theories at Low Energies}
In QCD the low-lying degrees of freedom are the pions, which according to Goldstone's Theorem result from the spontaneous breaking of chiral symmetry for $N_f \geq 2$. Since the mass of the pion is about $140$ MeV, pions are much lighter than the lightest non-Goldstone particles, such as the nucleons, which has a mass of about $\Lambda \sim 1$ GeV. Therefore, at sufficiently low energies the QCD partition function is well approximated by the partition function of an effective low-energy theory involving only pions. The QCD partition function in a finite Euclidian volume $V_4=L^4$ is dominated by the pions if
\begin{equation}
\frac{1}{\Lambda} \ll L.
\label{eq:domination1}
\end{equation}
This statement follows 
by comparing the contribution  of the pion, $\exp(-m_\pi L)$, to that of a heavier particle, $\exp(-\Lambda L)$. The interactions of the pions can be separated into zero-momentum modes and nonzero-momentum modes. It was realized by Gasser and Leutwyler that there exists a kinematic regime, where the fluctuations of the zero-momentum modes dominate the fluctuations of the nonzero-momentum modes \cite{Gasser:1987ah}. This regime is given by the condition
\begin{equation}
L \ll \frac{1}{m_\pi},
\label{eq:domination2}
\end{equation}
where $m_\pi$ is the pion mass. Intuitively this means that the wavelength of the pion is much larger than the linear extent of the box. Thus, the pion field does not vary appreciably over the size of the box, which result in small derivative terms. Therefore, it is possible to only consider the zero-momentum modes in the $\epsilon$-regime (\ref{eq:domination2}).
In this regime the Lie group can be factorized as $U(x)=U_0 \exp[i \sqrt2 \phi(x)/F_\pi]$, where $U_0$ is the zero-momentum part and $\phi(x)$ represents the fluctuation fields \cite{Damgaard:2006rh}. It turns out that to leading order one keeps only the static modes in the path integral, while the fluctuation degrees of freedom decouple. This can be seen as follows. When the isospin chemical potential couples to the theory it gives rise to a covariant time derivative in the effective $SU(2)$ Lagrangian \cite{Son:2000xc,Kogut:1999iv,Kogut:2000ek},

\begin{equation}
\partial_0 U(x) \rightarrow \nabla_0 U(x) = \partial_0 U(x) - \mu_{iso}[\sigma_3,U(x)],
\label{eq:covariant}
\end{equation}
where $\sigma_3$ is the usual Pauli matrix. The leading-order terms in the effective Lagrangian then read

\begin{eqnarray}
\mathcal{L} &=& \frac{F_\pi}{4} Tr[\nabla_0 U(x) \nabla_0 U^\dagger(x)+\partial_i U(x) \partial_i U^\dagger(x)] \nonumber\\
&& -\frac{\Sigma}{2} Tr[\mathcal{M} U^\dagger(x)+\mathcal{M}^\dagger U(x)],
\label{eq:chiral}
\end{eqnarray}
where $\mathcal{M}= {\rm diag}(m_u,m_d)$ is the quark mass matrix and $\Sigma$ is the chiral condensate. By expanding the Lie group

\begin{equation}
U(x)=U_0[1+i \sqrt2 \phi(x) /F_\pi+ \cdot \cdot \cdot],
\label{eq:expand}
\end{equation}
the usual kinetic term for $\phi(x)$ is produced. When power counting is used in the $\epsilon$-expansion it is assumed that  $m_\pi \sim p^2 = \mathcal{O}(\epsilon^2)$, while $\phi(x) \sim 1/L = \mathcal(\epsilon)$ and that a consistent power counting for the $\mu_{iso}$-term is $\mu_{iso}=\mathcal{O}(\epsilon^2)$ \cite{Gasser:1986vb,Gasser:1987ah,Giusti:2008fz}. Indeed, when the covariant derivative Eq. (\ref{eq:covariant}) is expanded using Eq. (\ref{eq:expand}), the leading contribution becomes

\begin{equation}
\nabla_0 U(x)=i\sqrt2 /F_\pi \partial_0 \phi(x) - \mu_{iso}[\sigma_3, U_0]+ \cdot \cdot \cdot.
\label{eq:Hej2}
\end{equation}
In the chiral Lagrangian Eq.(\ref{eq:chiral}) the mixed terms $\partial_0 \phi (x)[\sigma_3, U_0]$ produce only boundary contributions and play no role here. Thus, to leading order in the $\epsilon$-expansion the fluctuation field $\phi(x)$ gives rise only to the kinetic energy term \cite{Damgaard:2006rh}

\begin{equation}
\int d^4x \frac{1}{2} Tr \partial_\mu \phi(x) \partial_\mu \phi(x),
\label{eq:Hej3}
\end{equation}
which decouples as in the theory with $\mu_{iso}=0$.\\
Collecting the remaining terms it is seen that the leading contribution to the partition function in the $\epsilon$-regime is the zero-dimensional integral

\begin{widetext}
\begin{equation}
\mathcal{Z}^{N_f=2}(\mathcal{M}; \mu_{iso}) = \int_{SU(2)} dU e^{\frac{1}{4} V F_\pi^2 \mu_{iso}^2 Tr[U,\sigma_3][U^\dagger,\sigma_3]+\frac{1}{2}\Sigma VTr(\mathcal{M}^\dagger U +\mathcal{M} U^\dagger)},
\label{eq:Hej4}
\end{equation}
where the $0$-suffix on the group element $U \in SU(2)$ has been dropped for convenience.\\
Projection onto fixed gauge field topology $\nu$ is done by a Fourier transform, and amounts to the simple modification
\cite{Gasser:1986vb,Gasser:1987ah} 
\begin{equation}
\mathcal{Z}^{N_f=2}_\nu(\mathcal{M}; \mu_{iso}) = \int_{U(2)} dU (\det U)^\nu e^{\frac{1}{4} V F_\pi^2 \mu_{iso}^2 Tr[U,\sigma_3][U^\dagger,\sigma_3]+\frac{1}{2}\Sigma VTr(\mathcal{M}^\dagger U +\mathcal{M} U^\dagger)},
\label{eq:Eff.Part}
\end{equation}
where the leading-order contribution to the $\epsilon$-regime depends only on the scaling variables
\begin{equation}
\hat{m}_i \equiv m_i \Sigma V, \quad \quad \hat{\mu}_{iso}^2 \equiv \mu_{iso}^2 F_\pi^2 V.
\label{eq:scaling Variables}
\end{equation}
Derivation of the integral Eq. (\ref{eq:Eff.Part}) gives the following expression \cite{Splittorff:2003cu}

\begin{equation}
\mathcal{Z}^{N_f=2}_\nu (\hat{m}_u, \hat{m}_d; \hat{\mu}_{iso})
= 2 e^{2 \hat{\mu}_{iso}^2} \int_0^1 dt\ t\ e^{-2 \hat{\mu}_{iso}^2 t^2}
I_\nu(t\ \hat{m}_u) I_\nu(t\ \hat{m}_d).
\label{eq:Main}
\end{equation}
In sectors of fixed topological index $\nu$ the chiral condensate follows by differentiation of Eq. (\ref{eq:Main})
\begin{equation}
\langle \bar{\psi}\psi \rangle_\nu = \frac{1}{2} \partial_{\hat{m}} \log \mathcal{Z}^{N_f=2}_\nu(\hat{m};\hat{\mu}_{iso}) =\frac{\int_0^1 dt\ e^{-2 \hat{\mu}_{iso}^2 t^2}\ t^2\ I_{\nu}(\hat{m}t)(I_{\nu+1}(\hat{m}t)+I_{\nu-1}(\hat{m}t))}{\int_0^1 dt\ t\  e^{-2 \hat{\mu}_{iso}^2 t^2} I_\nu(\hat{m}t)^2},
\label{eq:HovedFormel}
\end{equation}
where for simplicity $\hat{m}=\hat{m}_u=\hat{m}_d$.\\
By summing over topology Eq. (\ref{eq:HovedFormel}) gives
\begin{eqnarray}
\langle \bar{\psi}\psi \rangle &=& \frac{\int_0^1 dt\ e^{-2 \hat{\mu}_{iso}^2 t^2}\ t^2\ \sum_{\nu= - \infty}^{\nu=\infty} I_{\nu}(\hat{m}t) (I_{\nu+1}(\hat{m}t)+I_{\nu-1}(\hat{m}t))}{\int_0^1 dt\ t\  e^{-2 \hat{\mu}_{iso}^2 t^2} \sum_{\nu= - \infty}^{\nu=\infty} (I_\nu(\hat{m}t)^2)} \nonumber \\
&=&\frac{\int_0^1 dt\ e^{-2 \hat{\mu}_{iso}^2 t^2}\ t^2\ (I_1(2\hat{m}t)+I_{-1}(2\hat{m}t))}{\int_0^1 dt\ t\  e^{-2 \hat{\mu}_{iso}^2 t^2} I_0(2\hat{m}t)}.
\label{eq:Sum over nu21}
\end{eqnarray}
For simplicity the vacuum angle $\theta$ has been set equal to zero \cite{Akemann:2001ir,Damgaard:1999ij}.

\end{widetext}

\begin{figure}[htb]
\epsfxsize=2.5in
\epsffile{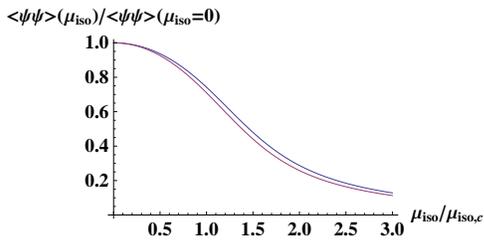}
\epsfxsize=2.5in
\vspace{0.1in}
\caption{\textit{\label{fig:Sum} The chiral condensate is plotted as a function of the isospin chemical potential for $\hat{m} = 1$. The upper line shows the full chiral condensate 
The lower line shows the case for $\nu=0$. It is seen how the simplified case, $\nu=0$, makes up a fine candidate for the full situation.}}
\end{figure}
In Fig.~\ref{fig:Sum} Eq. (\ref{eq:Sum over nu21}) is plotted as a function of $\mu_{iso}$. The upper line shows the case where a summation over $\nu$ has been made, and the bottom line shows the case where $\nu=0$. It is seen that the simplified case, $\nu=0$, does not diverge significantly from the full situation. Thus $\nu$ is set equal to zero from now on and the expression of the chiral condensate is given by

\begin{equation}
\langle \bar{\psi}\psi \rangle_{\nu=0} =\frac{2\int_0^1 dt\ e^{-2 \hat{\mu}_{iso}^2 t^2}\ t^2\ I_{0}(\hat{m}t)I_{1}(\hat{m}t)}{\int_0^1 dt\ t\  e^{-2 \hat{\mu}_{iso}^2 t^2} I_0(\hat{m}t)^2}.
\label{eq:Sum over nu2}
\end{equation}

\begin{figure}[htb]
\epsfxsize=1.6in
\epsffile{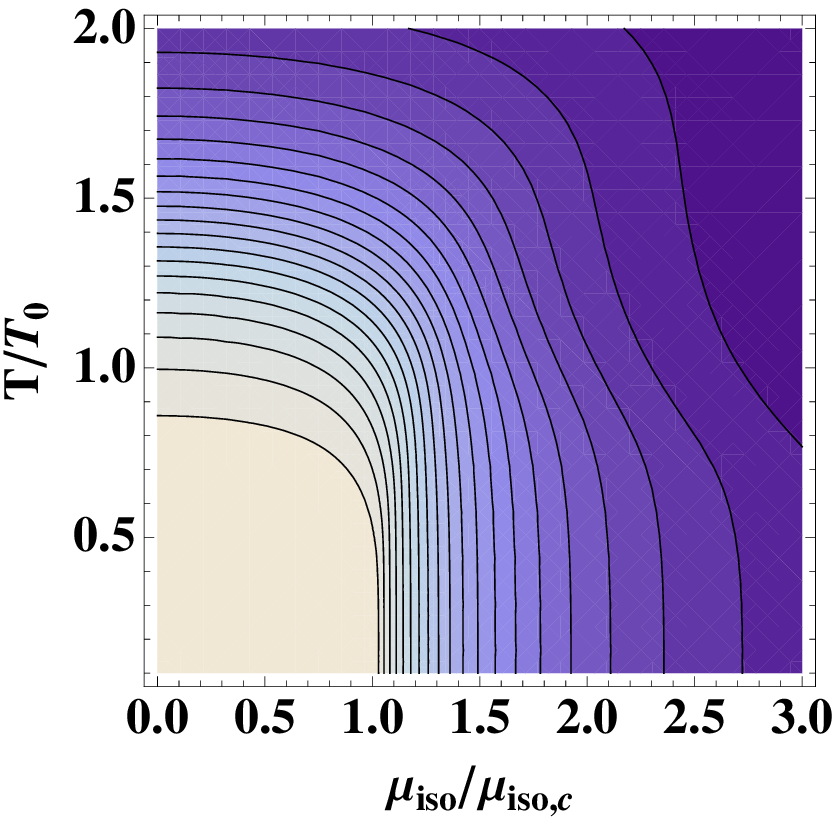}
\epsfxsize=1.6in
\epsffile{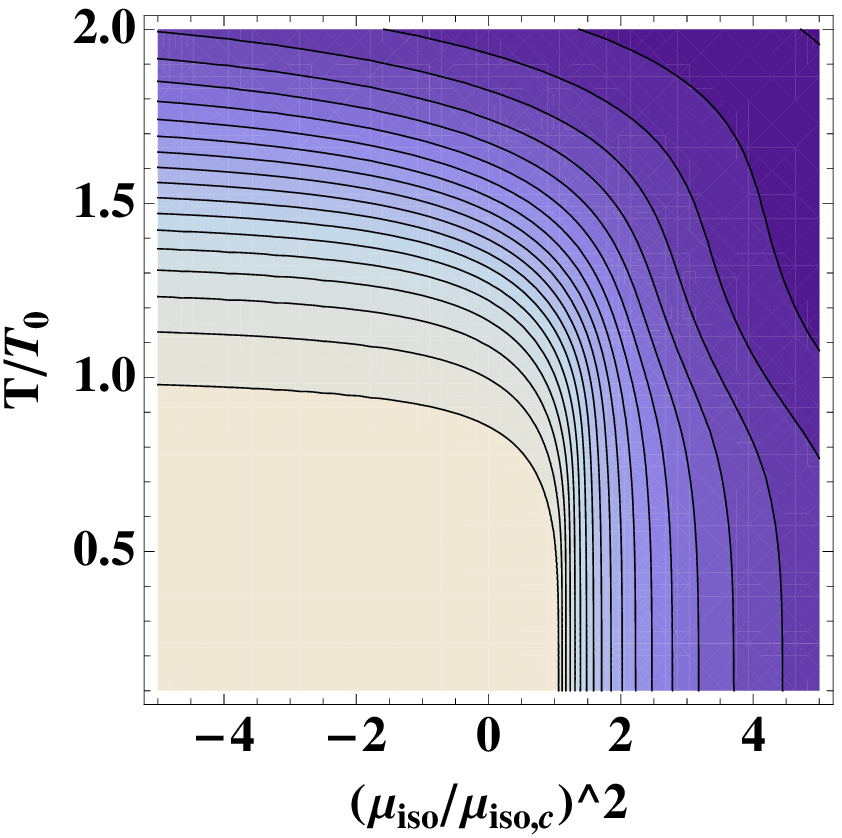}
\vspace{0.1in}
\epsfxsize=1.6in
\caption{\textit{\label{fig:ContourPlot} The figures show the evaluation of the chiral condensate in the $(\mu_{iso},T)$- and $(\mu_{iso}^2,T)$-plane respectively. The value of the condensate decrease with darker colours. To perform the plot one free parameter has to be chosen. In order to make the values on the axes dimensionless the free parameter has been picked to be $(2 \mu_c^2 F_\pi^2)/T^4_0= 10$, where $\mu_c =m_\pi/2$. It is emphasised that $T_0$ is not the critical temperature at which the QCD phase transition takes place, but is simply the scale chosen for these particularly plots.}}
\end{figure}
In Fig.~\ref{fig:ContourPlot} lines of constant values of the chiral condensate $\langle \bar{\psi}\psi \rangle$ are plotted in the $(\mu_{iso}, T)$- and in the $(\mu_{iso}^2,T)$-plane respectively. The increase in the imaginary plane is in agreement with \cite{cea-2009}.\\

The validity of the formulas fails at a temperature of $T \sim 150$ MeV. At this point the interactions among the massive particles becomes increasingly important. Thereby, the theory cannot be described by only considering the zero momentum modes in the $\epsilon$-regime.\\
By performing a three-loop analysis of the low temperature region of QCD, Gerber and Leutwyler have estimated the value of the critical temperature at which the chiral phase transition takes place \cite{Gerber:1988tt}. This is done by plotting the temperature dependence of the chiral condensate and following read off the temperature in the limit where 
$\langle \bar{\psi}\psi \rangle =0$. Their analysis indicates that in the chiral limit the phase transition occurs around
$$T_c \simeq 170\ {\rm MeV}.$$
This value is beyond the range of validity of the formulas, but since the order parameter falls rapidly at the upper end of the range, Gerber and Leutwyler consider it meaningful to make the estimate.\\ 
In this paper the critical temperature is estimated as the maximum value of the differentiated condensate with respect to temperature for fixed values of $\mu_{iso}$. 

\begin{figure}[htb]
\epsfxsize=2.5in
\epsffile{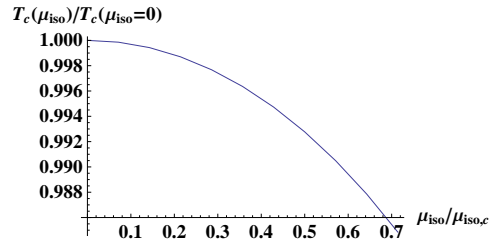}
\epsfxsize=2.5in
\vspace{0.1in}
\caption{\textit{\label{fig:Tc1} The critical temperature, $T_c$, is plotted as a function of the isospin chemical potential, $\mu_{iso}$, for $(2 \mu_c^2 F_\pi^2)/T^4_0=10$. The figure shows how $T_c$ slightly drops with increasing $\mu_{iso}$.}}
\end{figure}
In Fig.~\ref{fig:Tc1} the critical temperature is plotted as a function of $\mu_{iso}< m_\pi/2$. It is seen how $T_c$ slightly drops for increasing isospin chemical potential.\\

\noindent By Taylor expanding the critical temperature,
\begin{equation}
\frac{T_c(\mu_{iso})}{T_c(\mu_{iso}=0)}=1- A(\frac{\mu_{iso}}{\mu_{iso,c}})^2.
\end{equation}
A calculation of the average value of the curvature for $(2 \mu_c^2 F_\pi^2)/T^4_0=10$ gives $A \approx 0.01 $. This value is relatively stable for $(2 \mu_c^2 F_\pi^2)/T^4_0 \gtrsim 5$ and is comparable to the value obtained in \cite{kogut-2004-70} for the lowest quark mass.

\section{The Fermion Sign Problem \label{sign problem}}
As mentioned, the fermion determinant is complex at nonzero chemical potential. This prohibit standard Monte Carlo sampling of the path integral in what is known as the sign problem. Overviews with references to various studies of this problem may be found in \cite{deforcrand-2002-642,hands-2002-106,philipsen-2001-94,danzer-2009}.\\
\noindent
By factoring the determinant of the Dirac operator into its absolute value and the phase factor, $\exp(i\phi)$, the determinant can be rewritten as
\begin{equation}
\det(D+m+\mu \gamma_0)=|\det(D+m+\mu \gamma_0)|e^{i\phi}.
\end{equation}
The severity of the sign problem can be measured through the expectation value of the average phase factor. A physical interpretation is obtained by defining the phase with respect to the phase quenched partition function
\begin{equation}
\langle e^{2i\phi} \rangle_{pq}=\frac{\langle{\det}^2(D+m+\mu \gamma_0)\rangle}{\langle|\det(D+m+\mu \gamma_0)|^2\rangle} \equiv \frac{\mathcal{Z}_{QCD}^{N_f=2}}{\mathcal{Z}_{|QCD|}^{N_f=2}}.
\label{eq:Sign Problem3}
\end{equation}
By inserting Eq. (\ref{eq:Main}) in Eq. (\ref{eq:Sign Problem3}) the expectation value of the average phase factor becomes
\begin{equation}
\langle e^{2i\phi} \rangle_{pq} = \frac{\int_0^1 dt\ t\ I_\nu(t\ \hat{m})^2}{e^{2 \hat{\mu}_{iso}^2} \int_0^1 dt\ t\ e^{-2 \hat{\mu}_{iso}^2 t^2} I_\nu(t\ \hat{m})^2}.
\label{eq:Sign Problem2}
\end{equation}

\begin{figure}[htb]
\epsfxsize=1.7in
\epsffile{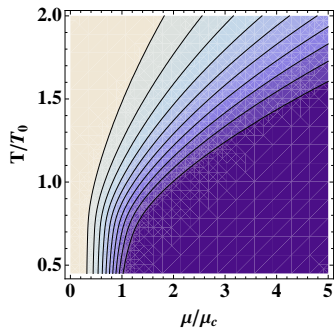}
\epsfxsize=1.7in
\vspace{0.1in}
\caption{\textit{\label{fig:SignProbNuLigNul}The expectation value of the average phase factor, $\langle e^{2i\phi} \rangle_{pq}$, is plotted in the $(\mu,T)$-plane for $(2 \mu_c^2 F_\pi^2)/T^4_0 = 1$. The value of $\langle e^{2i\phi} \rangle_{pq}$ increase with lighter colours. Since the fermion sign problem is mild, when the phase factor is of the order one, the figure shows how the sign problem is getting more mild with increasing temperature.}}
\end{figure}
According to \cite{splittorff-2006-023} the sign problem is mild if $\langle e^{2i\phi} \rangle_{pq}$ is of the order one.
In Fig.~\ref{fig:SignProbNuLigNul} Eq. (\ref{eq:Sign Problem2}) is plotted for $\nu=0$ in the $(\mu,T)$-plane. A plot after summation over $\nu$ has been made too, but since no significantly difference between the two plots were to be seen, only the case for $\nu=0$ is shown. Thus, the $\nu=0$ contribution makes up a fine candidate for the visualization of the sign problem.\\ 
The figure shows that the value of $\langle e^{2i\phi} \rangle_{pq}$ increase with larger temperature and thus makes the sign problem mild. Figure \ref{fig:SignProbNuLigNul} is in good agreement with \cite{Han:2008xj}, where random matrix theory (RMT) is applied to obtain an analytic expression for the average phase factor as a function of $T$, $\mu_B$ and $m$.

\section{Conclusions}
In this paper QCD has been considered in the $\epsilon$-regime of small isospin chemical potentials. Contrary to the regime of baryon chemical potentials, where the complex fermion determinant makes it impossible to directly simulate lattice QCD by probabilistic methods, the isospin chemical potentials makes up a very accommodating situation.\\
The $\epsilon$-regime was specified such that chiral perturbation theory was valid and that the Goldstone modes associated with chiral symmetry were the dominant degrees of freedom. Moreover, the Goldstone field was treated as constant. This had the advantage of allowing for exact analytic calculations.\\
Further, the effective theory has been used to derive an expression for the chiral condensate in the case of two quark flavors $N_f=2$, which has further been plotted in the $(\mu_{iso},T)$-plane and in the $(\mu_{iso}^2,T)$-plane. The lines of constant values of the chiral condensate showed a curvature in the real plane and an increase in the complex plane.\\ 
Additionally, an estimate of the critical temperature as a function of the isospin chemical potential has been given. The measured small decrease of the critical temperature proved to be comparable to the values obtained in lattice QCD \cite{kogut-2004-70}.\\
Finally, the fermion sign problem has been investigated by plotting the expectation value of the phase factor in the $(\mu,T)$-plane. It was seen how the phase factor gets larger with increasing temperature and thus makes the fermion sign problem more mild.

\begin{acknowledgments}
I wish to thank Poul Henrik Damgaard and Kim Splittorff for many useful discussions.
\end{acknowledgments}


\begin{thebibliography}{40}
\bibitem{Frontier} 
K.~Rajagopal and F.~Wilczek,
``At the Frontier of Particle Physics/Handbook of QCD'',
edited by M. Shifman (World Scientific, Singapore),
Vol.\ {\bf 3}, 2061 (2001).

\bibitem{hands-2002-106} 
  S.~Hands,
  Nucl.\ Phys.\ Proc.\ Suppl.\  {\bf 106}, 142 (2002)
  [arXiv:hep-lat/0109034].

\bibitem{Fodor:2001pe} 
Z.~Fodor and S.~D.~Katz,
  JHEP {\bf 0203}, 014 (2002)
  [arXiv:hep-lat/0106002].

\bibitem{deforcrand-2002-642} 
 P.~de Forcrand and O.~Philipsen,
  Nucl.\ Phys.\  B {\bf 642}, 290 (2002)
  [arXiv:hep-lat/0205016].

\bibitem{allton-2002-66} 
 C.~R.~Allton {\it et al.},
  Phys.\ Rev.\  D {\bf 66}, 074507 (2002)
  [arXiv:hep-lat/0204010].

\bibitem{Miyamura:2002en} 
O.~Miyamura, S.~Choe, Y.~Liu, T.~Takaishi and A.~Nakamura,
  Phys.\ Rev.\  D {\bf 66}, 077502 (2002)
  [arXiv:hep-lat/0204013].

\bibitem{Gavai:2002kq} 
R.~V.~Gavai and S.~Gupta,
  Phys.\ Rev.\  D {\bf 65}, 094515 (2002)
  [arXiv:hep-lat/0202006].

\bibitem{delia-2003-67} 
M.~D'Elia and M.~P.~Lombardo,
  Phys.\ Rev.\  D {\bf 67}, 014505 (2003)
  [arXiv:hep-lat/0209146].

\bibitem{splittorff-2006-023} 
K.~Splittorff,
  PoS {\bf LAT2006}, 023 (2006)
  [arXiv:hep-lat/0610072].

\bibitem{Rajagopal:2000wf}
  K.~Rajagopal and F.~Wilczek,
  [arXiv:hep-ph/0011333].

\bibitem{Verbaarschot:2000dy} 
J.~J.~M.~Verbaarschot and T.~Wettig,
  Ann.\ Rev.\ Nucl.\ Part.\ Sci.\  {\bf 50}, 343 (2000)
  [arXiv:hep-ph/0003017].

\bibitem{Gasser Leutwyler} 
J.~Gasser and H.~Leutwyler, 
Phys.\ Lett.\ B {\bf 188}, 477 (1987).

\bibitem{Gasser:1987zq} 
J.~Gasser and H.~Leutwyler,
Nucl.\ Phys.\ B {\bf 307}, 763 (1988).

\bibitem{Gasser:1986vb} 
J.~Gasser and H.~Leutwyler, 
Phys.\ Lett.\ B {\bf 184}, 83 (1987).

\bibitem{Alford:1998sd} 
M.~G.~Alford, A.~Kapustin and F.~Wilczek,
  Phys.\ Rev.\  D {\bf 59}, 054502 (1999)
  [arXiv:hep-lat/9807039].

\bibitem{Kogut:2002zg} 
J.~B.~Kogut and D.~K.~Sinclair,
  Phys.\ Rev.\  D {\bf 66}, 034505 (2002)
  [arXiv:hep-lat/0202028].

\bibitem{Kogut:2001id} 
J.~B.~Kogut and D.~Toublan,
  Phys.\ Rev.\  D {\bf 64}, 034007 (2001)
  [arXiv:hep-ph/0103271].

\bibitem{Kogut:2001if} 
J.~B.~Kogut, D.~Toublan and D.~K.~Sinclair,
  Phys.\ Lett.\  B {\bf 514}, 77 (2001)
  [arXiv:hep-lat/0104010].

\bibitem{Kogut:2002cm} 
J.~B.~Kogut, D.~Toublan and D.~K.~Sinclair,
  Nucl.\ Phys.\  B {\bf 642}, 181 (2002)
  [arXiv:hep-lat/0205019].

\bibitem{Leutwyler Smilga} 
H.~Leutwyler and A.~V.~Smilga,
Phys.\ Rev.\ D {\bf 46}, 5607-5632 (1992).

\bibitem{Klein:2003fy} 
B.~Klein, D.~Toublan and J.~J.~M.~Verbaarschot,
  Phys.\ Rev.\  D {\bf 68}, 014009 (2003)
  [arXiv:hep-ph/0301143].

\bibitem{Toublan:2004bj} 
D.~Toublan, B.~Klein and J.~J.~M.~Verbaarschot,
  Nucl.\ Phys.\ Proc.\ Suppl.\  {\bf 140}, 562 (2005)
  [arXiv:hep-lat/0409035].

\bibitem{deforcrand-2007-237} 
P.~de Forcrand, M.~A.~Stephanov and U.~Wenger,
  PoS {\bf LAT2007}, 237 (2007)
  [arXiv:0711.0023 [hep-lat]].

\bibitem{Halasz:1998qr} 
A.~M.~Halasz, A.~D.~Jackson, R.~E.~Shrock, M.~A.~Stephanov and J.~J.~M.~Verbaarschot,
  Phys.\ Rev.\  D {\bf 58}, 096007 (1998)
  [arXiv:hep-ph/9804290].

\bibitem{Splittorff:2001fy} 
K.~Splittorff, D.~Toublan and J.~J.~M.~Verbaarschot,
  Nucl.\ Phys.\  B {\bf 620}, 290 (2002)
  [arXiv:hep-ph/0108040].

\bibitem{Splittorff:2002xn} 
K.~Splittorff, D.~Toublan and J.~J.~M.~Verbaarschot,
  Nucl.\ Phys.\  B {\bf 639}, 524 (2002)
  [arXiv:hep-ph/0204076].

\bibitem{kogut-2004-70} 
J.~B.~Kogut and D.~K.~Sinclair,
  Phys.\ Rev.\  D {\bf 70}, 094501 (2004)
  [arXiv:hep-lat/0407027].

\bibitem{Son:2000xc} 
D.~T.~Son and M.~A.~Stephanov,
  Phys.\ Rev.\ Lett.\  {\bf 86}, 592 (2001)
  [arXiv:hep-ph/0005225].

\bibitem{Kogut:1999iv} 
J.~B.~Kogut, M.~A.~Stephanov and D.~Toublan,
  Phys.\ Lett.\  B {\bf 464}, 183 (1999)
  [arXiv:hep-ph/9906346].

\bibitem{Kogut:2000ek} 
J.~B.~Kogut, M.~A.~Stephanov, D.~Toublan, J.~J.~M.~Verbaarschot and A.~Zhitnitsky,
  Nucl.\ Phys.\  B {\bf 582}, 477 (2000)
  [arXiv:hep-ph/0001171].

\bibitem{Gasser:1987ah} 
J.~Gasser and H.~Leutwyler, 
Phys.\ Lett.\ B {\bf 188}, 477 (1987).

\bibitem{Giusti:2008fz}
  L.~Giusti, P.~Hernandez, S.~Necco, C.~Pena, J.~Wennekers and H.~Wittig,
  JHEP {\bf 0805}, 024 (2008)
  [arXiv:0803.2772 [hep-lat]].

\bibitem{Damgaard:2006rh} 
P.~H.~Damgaard, U.~M.~Heller, K.~Splittorff, B.~Svetitsky and D.~Toublan,
  Phys.\ Rev.\  D {\bf 73}, 105016 (2006)
  [arXiv:hep-th/0604054].

\bibitem{Splittorff:2003cu} 
K.~Splittorff and J.~J.~M.~Verbaarschot,
  Nucl.\ Phys.\  B {\bf 683}, 467 (2004)
  [arXiv:hep-th/0310271].

\bibitem{Akemann:2001ir} 
G.~Akemann, J.~T.~Lenaghan and K.~Splittorff,
  Phys.\ Rev.\  D {\bf 65}, 085015 (2002)
  [arXiv:hep-th/0110157].

\bibitem{Damgaard:1999ij} 
P.~H.~Damgaard,
  Nucl.\ Phys.\  B {\bf 556}, 327 (1999)
  [arXiv:hep-th/9903096].

\bibitem{cea-2009} 
P.~Cea, L.~Cosmai, M.~D'Elia, C.~Manneschi and A.~Papa,
  [arXiv:0905.1292 [hep-lat]].

\bibitem{Gerber:1988tt} 
P.~Gerber and H.~Leutwyler,
Nucl.\ Phys.\ B {\bf 321}, 387 (1989).

\bibitem{philipsen-2001-94} 
 O.~Philipsen,
  Nucl.\ Phys.\ Proc.\ Suppl.\  {\bf 94}, 49 (2001)
  [arXiv:hep-lat/0011019].

\bibitem{danzer-2009} 
 J.~Danzer, C.~Gattringer, L.~Liptak and M.~Marinkovic,
  [arXiv:0907.3084 [hep-lat]].

\bibitem{Han:2008xj} 
J.~Han and M.~A.~Stephanov,
  Phys.\ Rev.\  D {\bf 78}, 054507 (2008)
  [arXiv:0805.1939 [hep-lat]].

\end{thebibliography}
\end{document}